\documentclass{appolb}
\usepackage{epsfig}

\begin{document}
\newcommand{\ee}{${\rm e}^+{\rm e}^-$}

\title{Mixture of anisotropic fluids
\thanks{email addresses: wojciech.florkowski@ifj.edu.pl and radoslaw.maj@ujk.edu.pl}
}
\author{Wojciech Florkowski
\address{The H. Niewodnicza\'nski Institute of Nuclear Physics, Polish Academy of Sciences, PL-31342 Krak\'ow, Poland, and \\
Institute of Physics, Jan Kochanowski University, PL-25406~Kielce, Poland}
\and
Radoslaw Maj
\address{Institute of Physics, Jan Kochanowski University, PL-25406~Kielce, Poland}
}
\maketitle
\begin{abstract}
The recently introduced approach describing coupled quark and gluon anisotropic fluids is generalized to include explicitly the transitions between quarks and gluons. We study the effects of such processes on the thermalization rate of anisotropic systems. We find that the quark-gluon transitions may enhance the overall thermalization rate in the cases where the initial momentum anisotropies correspond to mixed oblate-prolate or prolate configurations. On the other hand, no effect on the thermalization rate is found in the case of oblate configurations. The observed regularities are connected with the late-time behavior of the analyzed systems which is described either by the exponential decay or the power law.
\end{abstract}

\vspace{1.0cm}
  
\section{Introduction}

The space-time dynamics of soft hadronic matter created in ultrarelativistic heavy-ion collisions is well described by relativistic viscous hydrodynamics \cite{Israel:1979wp,Muronga:2003ta,Baier:2006um,Romatschke:2007mq,Dusling:2007gi,Luzum:2008cw,Song:2008hj,Denicol:2010tr,Schenke:2011tv,Shen:2011eg,Bozek:2011wa,Niemi:2011ix,Bozek:2012qs}. Nevertheless, large gradients present in the early stages of the collisions imply that viscous corrections to the ideal energy momentum tensor are large and the system is highly anisotropic in the momentum space. Such large momentum-space anisotropies pose a problem for 2nd-order viscous hydrodynamics \cite{Martinez:2009mf}, since the latter relies on a linearization around an isotropic background. This has stimulated the development of reorganizations of viscous hydrodynamics in which one incorporates the possibility of large momentum-space anisotropies at leading order  \cite{Florkowski:2010cf,Martinez:2010sc,
Ryblewski:2010bs,Martinez:2010sd,Ryblewski:2011aq,
Martinez:2012tu,Ryblewski:2012rr}. 
The newly constructed framework has been dubbed {\it anisotropic hydrodynamics}. Recently, it has been demonstrated that for one-dimensional and boost-invariant systems the results of anisotropic hydrodynamics agree very well with the predictions of kinetic theory \cite{Florkowski:2013lza,Florkowski:2013lya}.

Anisotropic systems have been studied recently also within the AdS/CFT correspondence framework \cite{Mateos:2011ix,Mateos:2011tv,Chernicoff:2012iq,
Chernicoff:2012gu,Gahramanov:2012wz,
Kalaydzhyan:2013gja,Heller:2011ju,
Heller:2012je}. The general interest in anisotropic systems and the issue of fast thermalization of matter produced in heavy-ion collisions have motivated a recent study of a mixture of two (quark and gluon) anisotropic fluids \cite{Florkowski:2012as}. In this paper we generalize the approach introduced in Ref.~\cite{Florkowski:2012as} by including explicitly the transitions between quarks and gluons. We find that such processes may enhance the overall thermalization rate in the cases where the initial momentum anisotropies correspond to mixed oblate-prolate or prolate configurations. On the other hand, they do not enhance the thermalization rate in the case of oblate configurations. These regularities are connected with the time development of the systems which is described either by the exponential decay or by the power law.

\section{Kinetic equations}

The starting point for our analysis are the two coupled kinetic equations for quarks and gluons in the relaxation-time approximation
\begin{eqnarray}
p^\mu \partial_\mu Q &=& p \cdot U \left(
\frac{Q_{\rm eq}-Q}{\tau_{\rm eq}} 
+\frac{G}{\tau_g}
-\frac{Q}{\tau_q} 
\right), \label{kineqQ} \\
p^\mu \partial_\mu G &=& p \cdot U \left(
\frac{G_{\rm eq}-G}{\tau_{\rm eq}} 
+\frac{Q}{\tau_q}
-\frac{G}{\tau_g}
\right). \label{kineqG}
\end{eqnarray}
Here $Q(x,p)$ and $G(x,p)$ are quark and gluon phase-space distribution functions~\footnote{The quark distribution function $Q(x,p)$ includes in our approach both quarks and antiquarks.}, the parameter $\tau_{\rm eq}$ is the relaxation time characterizing a typical timescale for equilibration processes, and $U^\mu$ is the (hydrodynamic) flow of matter,
\begin{equation}
U^\mu = \gamma (1, v_x,v_y, v_z), \quad 
\gamma = (1-v^2)^{-1/2}.
\label{U}
\end{equation}

Compared to Ref.~\cite{Florkowski:2012as}, a novel feature of our approach is the introduction of extra terms on the right-hand sides of Eqs.~(\ref{kineqQ}) and (\ref{kineqG}). However, for sake of simplicity, we restrict our present study to the case where the relaxation times for quarks and gluons are the same and the baryon number of the system is zero. These two aspects have been analyzed in greater detail in Ref.~\cite{Florkowski:2012as}. The new terms describe the production of quarks (associated with the reduction of the gluon content in the system) and the production of gluons (associated with the reduction of the quark content in the system)~\footnote{A similar system of kinetic equations but in a different context was used in Ref.~\cite{Dolejsi:1994qp}.}. The production processes are characterized by the two timescale parameters (transition times) $\tau_q$ and $\tau_g$. We assume that the quark-gluon transition processes do not contribute to the thermalization processes characterized by the relaxation time $\tau_{\rm eq}$  --- the aim of the present study is to check whether the presence of additional processes may enhance the original thermalization rate. 

The phase-space distribution functions of quarks and gluons are normalized in the following way:
\begin{eqnarray}
g_q \int dP\, p^\mu Q(x,p) = N^\mu_q, \quad
g_g \int dP\, p^\mu G(x,p) = N^\mu_g,
\label{norm}
\end{eqnarray}
where $N^\mu_q$ and $N^\mu_g$ are particle number currents (of quarks and gluons, respectively), $g_q$ and $g_g$ are the degeneracy factors connected with internal degrees of freedom, and $dP=d^3p/((2\pi)^3E)$ is the Lorentz invariant momentum integration measure. In the numerical calculations we use the values $g_q = 24$ (a factor of 2 for antiparticles, 2 for spin, 2 for flavor, and 3 for color) and $g_g=16$ (a factor of 2 for spin and 8 for color) which lead to the ratio~\footnote{In Ref.~ \cite{Florkowski:2012as} quarks and antiquarks are treated separately and the degeneracy factor of quarks includes only spin, flavor, and color. This leads to the value $r=4/3$ used in Ref.~ \cite{Florkowski:2012as} which is consistent with the value $r=2/3$ used in this paper.}
\begin{eqnarray}
r = \frac{g_g}{g_q} = \frac{2}{3}.
\label{r}
\end{eqnarray}

In this work we neglect the effects of quantum statistics of particles. In equilibrium the quark and gluon distribution functions are equal and given by the Boltzmann distribution
\begin{eqnarray}
Q_{\rm eq} = G_{\rm eq} = \exp\left(
-\frac{p\cdot U}{T} \right),
\end{eqnarray} 
where $T$ is the system's temperature. The requirement that the equilibrium distributions with constant parameters become solutions of the kinetic equations (\ref{kineqQ}) and (\ref{kineqG}) leads to the detailed-balance condition
\begin{eqnarray}
\tau_q = \tau_g = \tau_{\rm tr}.
\label{detbal}
\end{eqnarray}
On the right-hand side of (\ref{detbal}) we have introduced the shorthand notation, $\tau_{\rm tr}$, for both $\tau_q$ and $\tau_g$.

\section{First and second moments of kinetic equation}

Integrating Eqs.~(\ref{kineqQ}) and (\ref{kineqG}) over three-momentum and multiplying each of them by the appropriate degeneracy factor yields
\begin{eqnarray}
\partial_\mu N^\mu_q &=& U_\mu \left(
\frac{N^\mu_{q, \rm eq}-N^\mu_q}{\tau_{\rm eq}} 
+\frac{1}{r}\frac{N^\mu_g}{\tau_{\rm tr}}
-\frac{N^\mu_q}{\tau_{\rm tr}} 
\right), \label{kineqQ1} \\
\partial_\mu N^\mu_g &=& U_\mu \left(
\frac{N^\mu_{g,\rm eq}-N^\mu_g}{\tau_{\rm eq}} 
+r\frac{N^\mu_q}{\tau_{\rm tr}}
-\frac{N^\mu_g}{\tau_{\rm tr}} 
\right). \label{kineqG1}
\end{eqnarray}
On the other hand, if we i) multiply Eqs.~(\ref{kineqQ}) and (\ref{kineqG}) by $p^\nu$, ii) integrate the two equations over three-momentum, iii) multiply each equation by the appropriate degeneracy factor, and  iv) add the two resulting expressions, then we find
\begin{eqnarray}
\partial_\mu T^{\mu\nu}_q 
+ \partial_\mu T^{\mu\nu}_g
= && U_\mu \left( 
\frac{T^{\mu\nu}_{q, \rm eq}
-T^{\mu\nu}_q}{\tau_{\rm eq}} 
+\frac{1}{r}\frac{T^{\mu\nu}_g}{\tau_{\rm tr}}
-\frac{T^{\mu\nu}_q}{\tau_{\rm tr}} 
\right) \nonumber \\
&& + \,\, U_\mu \left(\frac{T^{\mu\nu}_{g, \rm eq}
-T^{\mu\nu}_g}{\tau_{\rm eq}} 
+r \frac{T^{\mu\nu}_q}{\tau_{\rm tr}}
-\frac{T^{\mu\nu}_g}{\tau_{\rm tr}} 
\right). \label{kineqQG2} 
\end{eqnarray}
Here $T^{\mu\nu}_q $ and $T^{\mu\nu}_g$ are the energy-momentum tensors of quarks and gluons, respectively,
\begin{eqnarray}
g_q \int dP\, p^\mu p^\nu Q(x,p) = T^{\mu\nu}_q, \quad
g_g \int dP\, p^\mu p^\nu G(x,p) = T^{\mu\nu}_g.
\label{Tmunus}
\end{eqnarray}
It is important to notice that in order to have the energy-momentum conservation law, 
\begin{eqnarray}
\partial_\mu T^{\mu\nu}_q 
+ \partial_\mu T^{\mu\nu}_g =0,
\label{enmomcon}
\end{eqnarray} 
the right-hand side of Eq.~(\ref{kineqQG2}) should vanish. This requirement leads to so called Landau matching condition which is used to determine the effective temperature of the system~\footnote{The parameter $T$ in the equilibrium distributions has the meaning of the standard temperature only if the system is close or in equilibrium. Otherwise, $T$ serves as the measure of the energy density and may be treated as an effective temperature.}.

\section{Anisotropic-hydrodynamics framework}

Within the anisotropic-hydrodynamics framework one assumes that the distribution functions obtained from the kinetic equations are well approximated by the Romatschke-Strickland forms \cite{Romatschke:2003ms}
\begin{eqnarray}
Q(x,p) = \exp\left[-\frac{1}{\Lambda}
\sqrt{(p \cdot U)^2 + \xi_q (p \cdot V)^2} \right],
\nonumber \\
G(x,p) = \exp\left[-\frac{1}{\Lambda}
\sqrt{(p \cdot U)^2 + \xi_g (p \cdot V)^2} \right].
\label{RS}
\end{eqnarray}
Following \cite{Florkowski:2012as} we assume here that the quark and gluon distribution functions are characterized by the same transverse-momentum scale $\Lambda$, however, their anisotropy parameters $\xi_q$ and $\xi_g$ might be different.

The four-vector $V^{\mu}$ appearing in (\ref{RS}) defines the beam direction. It is defined by the formula
\begin{equation}
V^\mu = \gamma_z (v_z, 0, 0, 1), \quad \gamma_z = (1-v_z^2)^{-1/2}.
\label{V}
\end{equation}
The four-vectors $U^{\mu}$ and $V^{\mu}$ satisfy the following normalization conditions:
\begin{eqnarray}
U^2 = 1, \quad V^2 = -1, \quad U \cdot V = 0.
\label{UVnorm}
\end{eqnarray}
In the local-rest-frame of the fluid element one finds, 
\begin{eqnarray}
 U^\mu = (1,0,0,0), \quad V^\mu = (0,0,0,1). 
 \label{UVLRF}
\end{eqnarray}

The Lorentz structure of the Romatschke-Strickland form implies the following decomposition of the particle number currents and the energy-momentum tensors:
\begin{eqnarray}
N_i^\mu = n_i\, U^\mu, \quad (i = q,g),
\label{nqg}
\end{eqnarray}
\begin{eqnarray}
T_i^{\mu \nu} &=& \left( \varepsilon_i
+ P_{i \perp} \right)  U^\mu U^\nu
- P_{i \perp} g^{\mu\nu}
+\left( P_{i \parallel} - P_{i\perp} 
\right) V^\mu V^\nu 
\label{Tqg}
\end{eqnarray}
where
\begin{eqnarray}
n_i = \frac{g_i}{\pi^2} \frac{\Lambda^3}{\sqrt{1+\xi_i}}, 
\label{nqg1}
\end{eqnarray}
and
\begin{eqnarray}
\varepsilon_i = \frac{3 g_i \Lambda^4}{\pi^2} {\cal R}(\xi_i).
\label{epsqg}
\end{eqnarray}
The function ${\cal R}(\xi)$ appearing in (\ref{epsqg}) has the form \cite{Martinez:2010sc}
\begin{eqnarray}
{\cal R}(\xi) = \frac{1}{2(1+\xi)} \left[
1+ \frac{(1+\xi) \tan^{-1} \sqrt{\xi}}{\sqrt{\xi}}
\right].
\label{Rofxi}
\end{eqnarray}
Below we shall need also the expressions for the sum of the energy density and the longitudinal pressure (the longitudinal enthalpy)
\begin{eqnarray}
\varepsilon_i + P_{i\parallel}
= -\frac{6 g_i \Lambda^4}{\pi^2}(1+\xi_i) {\cal R}^\prime (\xi_i).
\label{entqg}
\end{eqnarray}

In equilibrium, the energy-momentum tensors have the well-known structure
\begin{eqnarray}
T^{\mu \nu}_{i, \rm eq} &=& 
\left( \varepsilon_{i, \rm eq}
+ P_{i, \rm eq} \right)  U^\mu U^\nu
- P_{i, \rm eq} g^{\mu\nu} ,
\label{TEQqg}
\end{eqnarray}
where
\begin{eqnarray}
n_{i, \rm eq} = \frac{g_i}{\pi^2} T^3,
\label{nqg1eq}
\end{eqnarray}
and
\begin{eqnarray}
\varepsilon_{i, \rm eq} = \frac{3 g_i T^4}{\pi^2},
\quad
P_{i, \rm eq} = \frac{\varepsilon_{i, \rm eq}}{3}.
\label{epseqqg}
\end{eqnarray}
We note that in Eqs.~(\ref{nqg})--(\ref{epsqg}) and (\ref{entqg})--(\ref{epseqqg}) the index $i$ takes the values $i=q$ for quarks and $i=g$ for gluons.

\section{Dynamic equations}
\label{sect:dyneq}

In this Section we derive four equations for the two anisotropy parameters, $\xi_q$ and $\xi_g$, the transverse-momentum scale, $\Lambda$, and the effective temperature, $T$, which can be used to determine the time dependence of these quantities in the case of one-dimensional and boost-invariant expansion. 

We start with the analysis of the zeroth moments of the kinetic equations. Substituting Eqs.~(\ref{nqg})  into Eqs.~(\ref{kineqQ1}) and (\ref{kineqG1}) we find
\begin{eqnarray}
\partial_\mu (n_q U^\mu) =
\frac{n_{q, \rm eq}-n_q}{\tau_{\rm eq}}
+ \frac{1}{r} \, \frac{n_g}{\tau_{\rm tr}} 
-\frac{n_q}{\tau_{\rm tr}}, 
\nonumber \\
\partial_\mu (n_g U^\mu) =
\frac{n_{g, \rm eq}-n_g}{\tau_{\rm eq}}
+ r \, \frac{n_q}{\tau_{\rm tr}} 
-\frac{n_g}{\tau_{\rm tr}} .
\label{eqs1a}
\end{eqnarray}
For one-dimensional and boost-invariant expansion one may check that \mbox{$\partial_\mu U^\mu = 1/\tau$} and $U^\mu \partial_\mu = d/d\tau$, where $\tau=\sqrt{t^2-z^2}$ is the longitudinal proper time. Hence, Eqs.~(\ref{eqs1a}) may be rewritten as
\begin{eqnarray}
\frac{d}{d\tau} \ln n_q + \frac{1}{\tau} =
\frac{1}{\tau_{\rm eq}} \left(
\frac{n_{q, \rm eq}}{n_q} -1 \right)
+ \frac{1}{\tau_{\rm tr}} \left(
\frac{1}{r} \, \frac{n_g}{n_q} - 1 \right),
\nonumber \\
\frac{d}{d\tau} \ln n_g + \frac{1}{\tau} =
\frac{1}{\tau_{\rm eq}} \left(
\frac{n_{g, \rm eq}}{n_g} -1 \right)
+ \frac{1}{\tau_{\rm tr}} \left(
r \, \frac{n_q}{n_g} -1 \right).
\label{eqs1b}
\end{eqnarray}
Using the expressions for the quark and gluon densities (\ref{nqg1}) in (\ref{eqs1b}) we find
\begin{eqnarray}
\frac{3}{\Lambda}\frac{d\Lambda}{ d\tau}
-\frac{1}{2(1+\xi_q)} \frac{d\xi_q}{d\tau}
+\frac{1}{\tau} = S_q
,
\nonumber \\ 
\frac{3}{\Lambda}\frac{d\Lambda}{ d\tau}
-\frac{1}{2(1+\xi_g)} \frac{d\xi_g}{d\tau}
+\frac{1}{\tau} = S_g,
\nonumber \\ \label{eq12}
\end{eqnarray}
where the right-hand sides are given by the formulas
\begin{eqnarray}
S_q = \frac{\kappa_q -1}{\tau_{\rm eq}} 
+ \frac{r_q -1}{\tau_{\rm tr}},
\nonumber \\
S_g = \frac{\kappa_g -1}{\tau_{\rm eq}} 
+ \frac{r_g -1}{\tau_{\rm tr}}, 
\end{eqnarray}
with 
\begin{eqnarray}
\kappa_i = \frac{T^3}{\Lambda^3}
\sqrt{1+\xi_i} \quad (i=q,g), \quad 
r_q = \frac{\sqrt{1+\xi_q}}{\sqrt{1+\xi_g}}, \quad
r_g = \frac{\sqrt{1+\xi_g}}{\sqrt{1+\xi_q}}.
\label{kappasr}
\end{eqnarray}

In the next step we turn to discussion of the first moment of the kinetic equation. The condition that the right-hand side of (\ref{kineqQG2}) is zero leads us to the constraint for the energy-densities
\begin{eqnarray}
\varepsilon_{q, \rm eq} + 
\varepsilon_{g, \rm eq} = 
\varepsilon_q \left(1 + \frac{\tau_{\rm eq}}{\tau_{\rm tr}} \left(1-r\right) \right) +
\varepsilon_g \left(1 + \frac{\tau_{\rm eq}}{\tau_{\rm tr}} \left(1-\frac{1}{r}\right) \right).
\label{LM1}
\end{eqnarray}
In the limit $\tau_{\rm tr} \to \infty$, Eq. ~(\ref{LM1}) is reduced to the standard Landau matching condition stating that the energy density of the system, $\varepsilon = \varepsilon_q+\varepsilon_g$, is equal to the energy density obtained from the thermal (background) distributions, $\varepsilon_{\rm eq} = \varepsilon_{q, \rm eq}+\varepsilon_{g, \rm eq}$. We note that Eq.~(\ref{LM1}) reproduces the standard Landau matching condition also in the case $r=1$. 

Since $r < 1$ in our case, the second term on the right-hand side of Eq.~(\ref{LM1}) may be negative and, in special cases, the whole right-hand side of  this equation may become smaller than zero (in the latter case Eq. ~(\ref{LM1}) has no solutions since its left-hand side is always positive). To avoid such situations, we impose the following condition on the relaxation and transition times
\begin{eqnarray}
\tau_{\rm eq} < \frac{r}{1-r} \,\, \tau_{\rm tr}.
\label{teqtan}
\end{eqnarray}
Clearly, this constraint reflects limitations of our simple kinetic model. Using Eqs.~(\ref{epsqg}) and (\ref{epseqqg}) in (\ref{LM1}) one finds
\begin{eqnarray}
T^4 = \Lambda^4 \left[
w_q {\cal R}(\xi_q) + w_g  {\cal R}(\xi_g) \right],
\label{eq3}
\end{eqnarray}
where the coefficients $w_q$ and $w_g$ (satisfying the normalization condition $w_q+w_g=1$) are defined by the expressions
\begin{eqnarray}
w_q &=& \frac{1 + \delta \left(1-r\right)}{1+r}, 
\nonumber \\
w_g &=&  \frac{r \left(1 + \delta \left(1-\frac{1}{r}\right) \right)}{1+r}.
\label{ab}
\end{eqnarray}
On the right-hand sides of Eqs.~(\ref{ab}) we have introduced the ratio of the equilibration and transition times
\begin{eqnarray}
\delta = \frac{\tau_{\rm eq}}{\tau_{\rm tr}}.
\label{delta}
\end{eqnarray}
The condition (\ref{teqtan}) implies that $\delta$ should satisfy the inequality $0 \leq \delta \leq 2$. 

We note that Eq.~(\ref{eq3}) allows us to calculate the ratio $T^3/\Lambda^2$ needed in (\ref{eq12}). The last required equation is obtained from the energy-momentum conservation law (\ref{enmomcon}) which is reduced to the single equation, namely
\begin{eqnarray}
\frac{d}{d\tau} \left(
\varepsilon_q+\varepsilon_g \right)
= -\frac{1}{\tau} \left(
\varepsilon_q + P_{q \parallel} +\varepsilon_g
+ P_{g \parallel} \right).
\label{enmomcon1}
\end{eqnarray}
Using Eqs.~(\ref{epsqg}) and (\ref{entqg}) one may rewrite Eq.~(\ref{enmomcon1}) in the form
\begin{eqnarray}
\frac{d}{d\tau} \left[\Lambda^4 \left(
{\cal R}(\xi_q) + r {\cal R}(\xi_g) \right) \right]
= \frac{2 \Lambda^4}{\tau} \left[
(1+\xi_q) {\cal R}^\prime (\xi_q) + 
r (1+\xi_g) {\cal R}^\prime (\xi_g) \right]. \quad
\label{eq4}
\end{eqnarray}

\begin{figure}[t]
\begin{center}
\includegraphics[angle=0,width=0.8\textwidth]{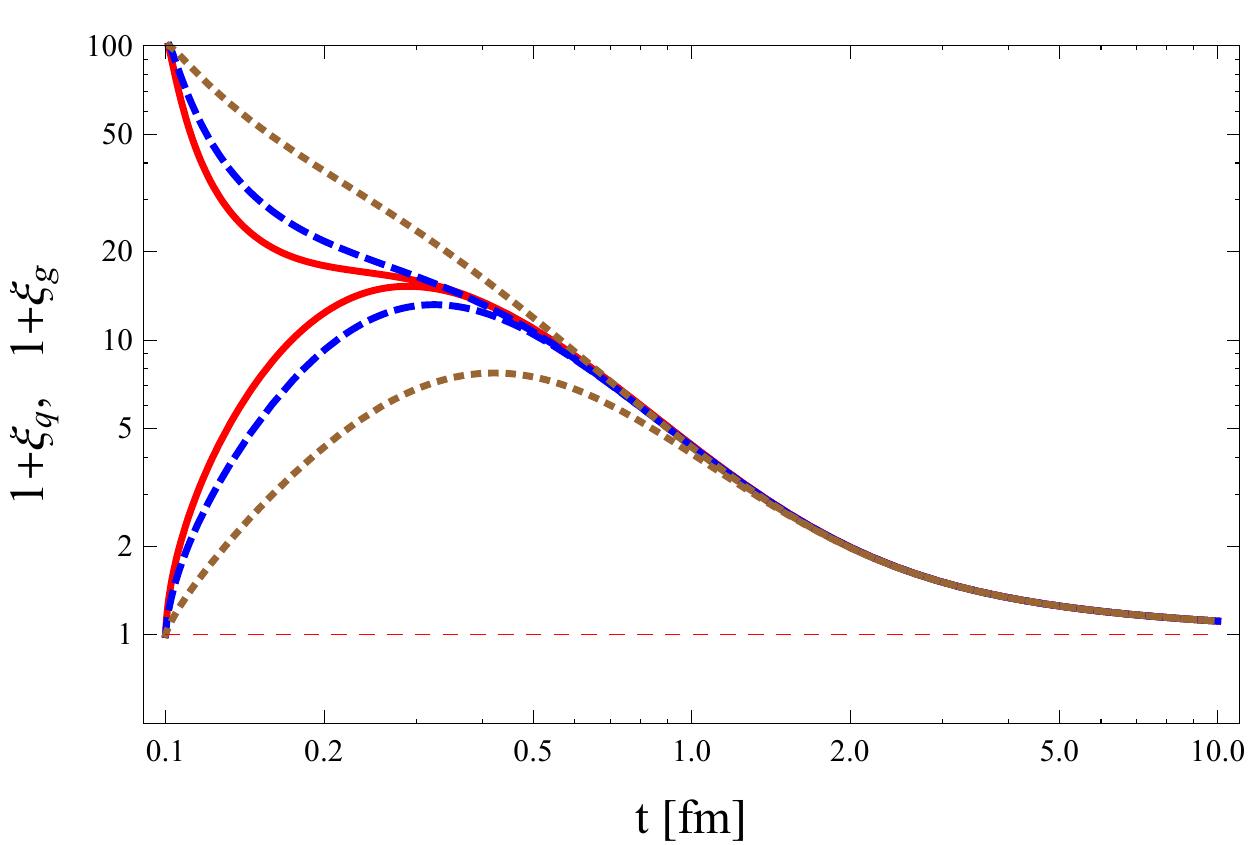} 
\end{center}
\caption{(Color online) The time dependence of the anisotropy parameters $\xi_g$ (three upper curves) and $\xi_q$ (three lower curves) for the initial conditions $\xi_g(\tau_0)=100$ and $\xi_q(\tau_0)=0$, and with the initial time $\tau_0=0.1$ fm/c. The initial momentum anisotropy is oblate for gluons and isotropic for quarks (an overall oblate configuration). The solid, dashed, and dotted curves correspond to the three cases: $\delta = 2,  1$ and $0$, respectively.}
\label{fig:xi1}
\end{figure}

\section{Results}
\label{sect:results}

In this Section we present our numerical solutions of Eqs.~(\ref{eq12}), (\ref{eq3}), and (\ref{eq4}). To allow for easy comparisons with earlier studies we use the same initial conditions as those used in Ref.~\cite{Florkowski:2012as}.

In Fig.~\ref{fig:xi1} we show the proper-time dependence of the anisotropy parameters $\xi_g$ (three upper curves) and $\xi_q$ (three lower curves) obtained with the initial conditions $\xi_g(\tau_0)=100$ and $\xi_q(\tau_0)=0$, and with the initial time $\tau_0=0.1$ fm/c. The initial momentum anisotropy is oblate for gluons and isotropic for quarks (this may be regarded as an overall oblate configuration). The solid, dashed, and dotted curves correspond to the three cases: $\delta = 2,  1$ and $0$, respectively. 

\begin{figure}[t]
\begin{center}
\includegraphics[angle=0,width=0.8\textwidth]{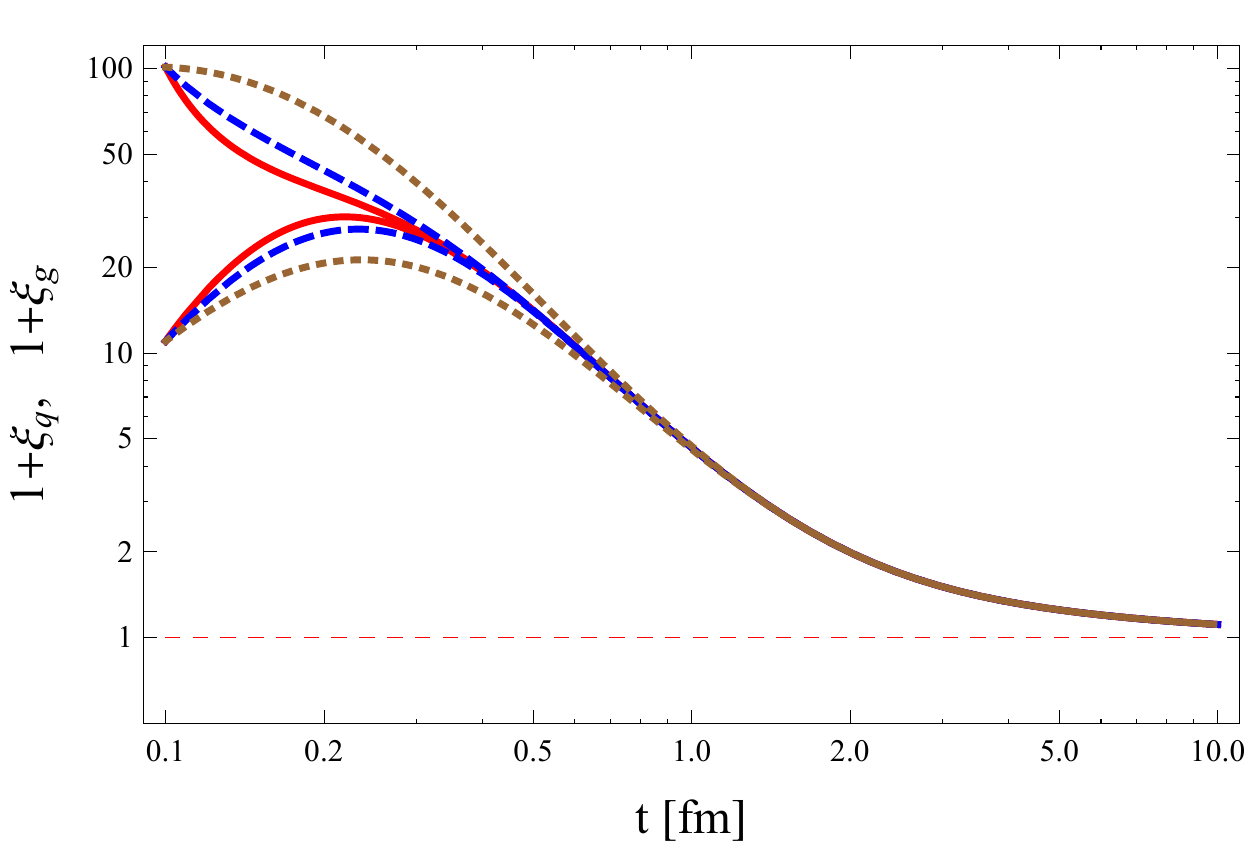} 
\end{center}
\caption{(Color online) The same as Fig.~\ref{fig:xi1} but with the initial anisotropies \mbox{$\xi_g(\tau_0)=100$} and $\xi_q(\tau_0)=10$. The initial momentum anisotropy is oblate for both gluons and quarks. }
\label{fig:xi2}
\end{figure}

The case $\delta=0 \,\,\, (\tau_{\rm tr} \to \infty)$ describes the situation analyzed previously in Ref.~\cite{Florkowski:2012as}. We observe that the presence of the quark-gluon transitions speeds up the convergence of the two anisotropies to each other --- if $\delta > 0$ they become equal within a shorter time interval than in the case $\delta = 0$. On the other hand, the time needed for the overall thermalization of the system (defined as the time when $\xi_q \approx \xi_g \approx 0$) is not changed. Similar situation is shown in Fig.~\ref{fig:xi2} where the calculations are done with the initial conditions $\xi_g(\tau_0)=100$ and $\xi_q(\tau_0)=10$. In this case the initial momentum anisotropy is oblate for both gluons and quarks.

\begin{figure}[t]
\begin{center}
\includegraphics[angle=0,width=0.8\textwidth]{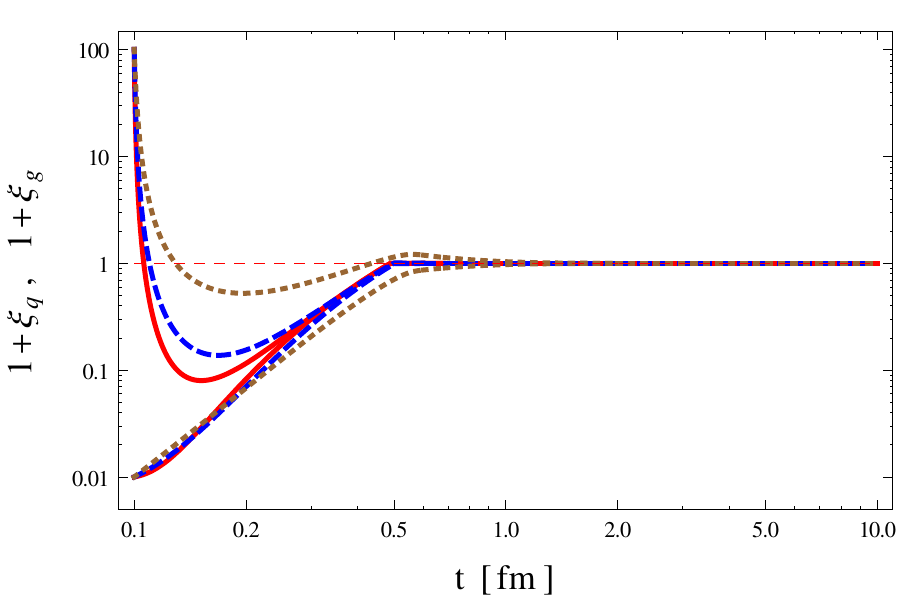} 
\end{center}
\caption{(Color online) The same as Fig.~\ref{fig:xi1} but with $\xi_g(\tau_0)=100$ and $\xi_q(\tau_0)=-0.99$. The initial momentum anisotropy is oblate for gluons and prolate for quarks. }
\label{fig:xi3}
\end{figure}

\begin{figure}[t]
\begin{center}
\includegraphics[angle=0,width=0.8\textwidth]{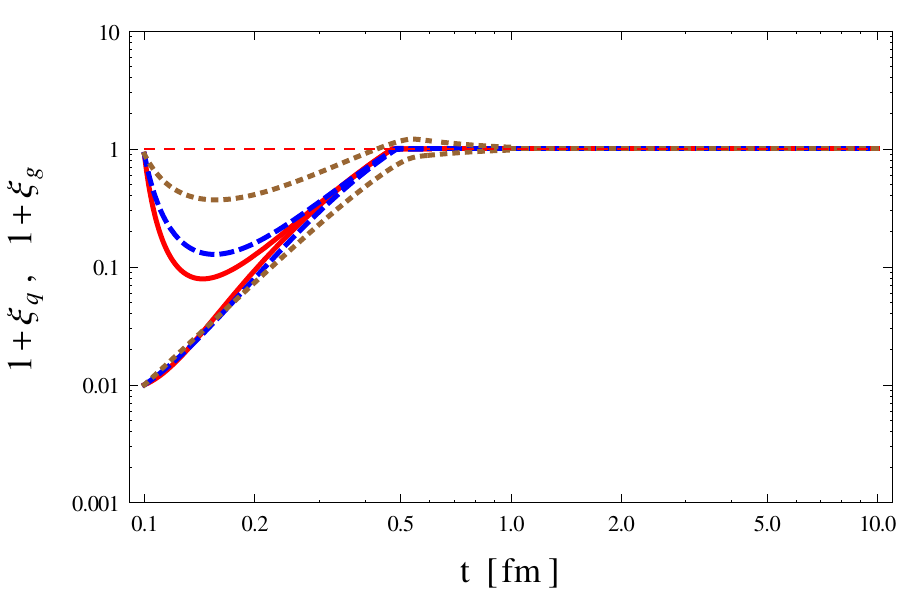} 
\end{center}
\caption{(Color online) The same as Fig.~\ref{fig:xi1} but with $\xi_g(\tau_0)=-0.10$ and $\xi_q(\tau_0)=-0.99$. The initial momentum anisotropy is prolate for both gluons and quarks.  }
\label{fig:xi4}
\end{figure}

A new qualitative behavior is shown in Fig.~\ref{fig:xi3} where the calculations are done with the initial conditions $\xi_g(\tau_0)=100$ and $\xi_q(\tau_0)=-0.99$. This situation corresponds to the initial momentum anisotropy which is oblate for gluons and prolate for quarks. The results shown in Fig.~\ref{fig:xi3} indicate that the transition processes increase the convergence of the two anisotropies to each other and speed up the overall thermalization time. Similar behavior is shown in Fig.~\ref{fig:xi4} where the initial anisotropies are $\xi_g(\tau_0)=-0.10$ and $\xi_q(\tau_0)=-0.99$. In this case, the initial momentum anisotropy is prolate for both gluons and quarks.

\section{Conclusions}
\label{sect:conc}

In order to conclude our study we comment in more detail on the connection between initial configurations and the role of the transition processes. It has been found already in \cite{Florkowski:2012as} that the late-time behavior is characterized by the exponential decay (for prolate or mixed oblate-prolate configurations) or power law (for oblate configurations). Below, in Sect. \ref{sect:ltb}, we give analytic arguments that the presence of transition processes does not change the qualitative character of the solutions --- we again deal either with the exponential decay or the power law. However, our new finding is that the exponential decay is characterized by the timescale parameter $\tau_{\rm eq}/(1+2\delta)$. This leads to a shorter overall thermalization rate in the prolate and oblate-prolate cases with the quark-gluon transitions included (where $\delta > 0$). We must emphasize, however, that all exponential solutions are characterized by very short timescales. Much larger differences exist between the exponential and power-law cases.

Finally, we want to make a remark that the appearance of the exponential solutions in the highly non-linear system of equations might be directly connected with the fast thermalization problem but this issue deserves a separate and more general study.

\medskip
This work was supported in part by the Polish National Science Center with Decision No. DEC-2012/06/A/ST2/00390.

\section{Appendix: Late-time behavior}
\label{sect:ltb}

Equations (\ref{eq12}), (\ref{eq3}), and (\ref{eq4}) form a closed system which can be used to determine the time evolution of four parameters: $\xi_q$, $\xi_g$, $\Lambda$, and  $T$. We may reduce this system of equations to two equations if we determine the ratio $T/\Lambda$ from Eq.~(\ref{eq3}) and calculate the proper-time derivative $d\Lambda/d\tau$ from (\ref{eq4}) and substitute these two quantities into Eq.~(\ref{eq12}). In this way we obtain two coupled differential equations for the anisotropy parameters
\begin{eqnarray}
\Delta(\xi_q,\xi_g) \, \left[\frac{d\xi_q}{2(1+\xi_q)  d\tau} - \frac{1}{\tau} \right] &=&  
P_q(\xi_q,\xi_g),
\label{EQ1q1a}
\end{eqnarray}
\begin{eqnarray}
\Delta(\xi_q,\xi_g) \, \left[\frac{d\xi_g}{2(1+\xi_g) d\tau} - \frac{1}{\tau} \right] &=&    
P_g(\xi_q,\xi_g).
\label{EQ1g1a}
\end{eqnarray}
Here
\begin{eqnarray}
\Delta = 4 \left[{\cal R}(\xi_q) +  r  {\cal R}(\xi_g)\right] + 
  6 \left[(1 + \xi_q)  {\cal R}^\prime(\xi_q) + 
  r (1 + \xi_g) {\cal R}^\prime(\xi_g) \right],
\label{Delta}
\end{eqnarray}
\begin{eqnarray}
P_q =  6 r (S_g - S_q) (1 + \xi_g)  {\cal R}^\prime(\xi_g) - 4 S_q \left[ R(\xi_q) + r  R(\xi_g) \right],
\label{Pq}
\end{eqnarray}
and
\begin{eqnarray}
P_g =  6 (S_q - S_g) (1 + \xi_q)  {\cal R}^\prime(\xi_q) - 4 S_g \left[ R(\xi_q) + r  R(\xi_g) \right].
\label{Pg}
\end{eqnarray}

Equations~(\ref{EQ1q1a}) and (\ref{EQ1g1a}) are convenient to analyze the late-time behavior of the anisotropy parameters. As the system approaches thermal equilibrium, $\xi_q$ and $\xi_g$ tend to zero. Hence, to study the late-time behavior of the system we include the terms up to {\it second order} in $\xi_q$ and $\xi_g$ on both sides of (\ref{EQ1q1a}) and (\ref{EQ1g1a}). The expansion of the function $\Delta$ starts with the linear terms,
\begin{equation}
\Delta = \frac{8}{15} \left( \xi_q + r \xi_g \right) 
- \frac{24}{35} \left(\xi_q^2 + r \xi_g^2 \right) + \ldots \,,
\label{Delta-app}
\end{equation}
therefore, we neglect $\xi_q$ and $\xi_g$ in the terms $(1+\xi_q)$ and $(1+\xi_g)$.

Instead of using (\ref{EQ1q1a}) and (\ref{EQ1g1a}) directly, we consider now their difference and the sum. For the difference we obtain
\begin{eqnarray}
\Delta(\xi_q,\xi_g) \, \left[\frac{d\xi_q}{2\,d\tau} - \frac{d\xi_g}{2\,d\tau} 
+\frac{\kappa_q-\kappa_g}{\tau_{\rm eq}}
+\frac{r_q-r_g}{\tau_{\rm tr}}
\right] &=& 0.
\label{diff}
\end{eqnarray}
This equation is fulfilled up to the second order if the equation in the square brackets is fulfilled in the first order. This leads to the equation
\begin{equation}
\frac{d}{d\tau} (\xi_q - \xi_g) = - \frac{1+ 2\delta}{\tau_{\rm eq}} (\xi_q - \xi_g) \equiv
-\, \alpha \, (\xi_q - \xi_g) ,
\label{deltadot}
\end{equation}
which for a constant relaxation time has the simple exponential solution 
\begin{equation}
d \equiv  \xi_q - \xi_g = d_0 \exp(-\alpha \tau),
\label{diffsol}
\end{equation}
where $d_0$ is an integration constant. Another useful quantity to consider is the linear combination of the two anisotropies
\begin{equation}
\xi \equiv \frac{\xi_q + r \, \xi_g}{1+r} .
\label{sumsol}
\end{equation}
Adding Eqs.~(\ref{EQ1q1a}) and (\ref{EQ1g1a}) multiplied by 1 and $r$, respectively, and keeping again all the terms up to second order gives 

\begin{eqnarray}
 \frac{\Delta^{(1)}}{2}  \frac{d\xi}{d\tau} - \frac{\Delta^{(1)}+\Delta^{(2)}}{\tau}  &=&  
P^{(2)},
\label{EQ}
\end{eqnarray}
where $P^{(2)}$ includes the second-order terms of the quantity 
\begin{equation}
P = \frac{P_q + r P_g}{(1+r)}.
\label{p}
\end{equation}
Below we show that the linear terms are absent in the expansion of $P$. 

Similarly, $\Delta^{(1)}$ and $\Delta^{(2)}$ denote the linear and quadratic terms in the expansion of $\Delta$. Straightforward calculations yield (below, for simplicity of notation, we present the results obtained with $r=2/3$)
\begin{equation}
\Delta^{(1)} = \frac{8 }{9}\,\xi\,,
\label{Delta1}
\end{equation}
\begin{equation}
\Delta^{(2)} = -\frac{8}{7} \,  \xi^2
- \frac{48}{175} \, d^2 ,
\label{Delta2}
\end{equation}
and 
\begin{equation}
P^{(2)} = \frac{1}{\tau_{\rm eq}} \left(
\frac{54+225\,\delta+5 \delta^2}{450} \,d^2
-\frac{4\delta}{45}\,  \xi \,d -\frac{2}{9} \, \xi^2
\right).
\label{p2}
\end{equation}

In Eq.~(\ref{EQ}) we may drop the term $\Delta^{(2)}/\tau$, since $\Delta^{(1)}/\tau$ is the leading term in the  whole expression. In this way, we obtain
\begin{equation}
\xi \left( \frac{d \xi}{d\tau} - \frac{2}{\tau} \right)
=\frac{1}{\tau_{\rm eq}}
\left[a_1 \,d^2 +b_1 \, \xi \,d +c_1\, \xi^2 \right],
\label{apEQ}
\end{equation}
where $a_1 = (54+225\,\delta+5 \delta^2)/200$, $b_1=-\delta/5$, and $c_1=-1/2$.

In the next step, we search for the solutions of Eq.~(\ref{apEQ}) in the form \mbox{$\xi(\tau) = z(\tau) e^{-\alpha\tau}$}. It is also convenient to introduce a dimensionless parameter $s=\alpha \tau$. In this way we obtain
\begin{equation}
\frac{dz}{ds} = \frac{2 e^{s}}{s} + 
\frac{a + b z + c z^2}{z},
\label{dzds}
\end{equation}
where $a=a_1 d_0^2/(1+2\delta)$, $b=b_1/(1+2\delta) d_0$, and $c=c_1/(1+2\delta)+1$. One may check that the numerator of the last fraction on the right-hand side of Eq.~(\ref{dzds}) is always positive. Hence, if $z$ is positive at $s=s_0$ it will further increase with $s$. On the other hand, if $z$ is negative at $s=s_0$, the competition of the two terms on the right-hand side of Eq.~(\ref{dzds}) forces $z$ to approach zero.

To see this behavior in more detail we observe that Eq.~(\ref{dzds}) has two characteristic asymptotic  solutions. In the first case $z$ is very small and negative, so the terms $b z$ and $c z^2$ may be neglected compared to $a$. The approximate equation
\begin{equation}
\frac{dz}{ds} = \frac{2 e^{s}}{s} + 
\frac{a}{z},
\label{dzds1}
\end{equation}
has the asymptotic solution of the form
\begin{eqnarray}
z(s) = -\frac{a s}{2} e^{-s},
\label{z1}
\end{eqnarray}
which supports our assumptions leading from (\ref{dzds}) to (\ref{dzds1}). In the second case $z$ is positive and grows with $s$, hence Eq.~(\ref{dzds}) may be approximated by the equation
\begin{equation}
\frac{dz}{ds} = \frac{2 e^{s}}{s} + 
c z,
\label{dzds2}
\end{equation}
whose solution is
\begin{eqnarray}
z(s) = C e^{c s} + 2 e^{c s} \hbox{Ei}\left((1-c) s\right).
\label{z2}
\end{eqnarray}
where $C$ is an integration constant. Using the relation $\xi = z e^{-s}$ we find
\begin{eqnarray}
\xi(s) = C e^{(c-1) s} + 2 e^{(c-1) s} \hbox{Ei}\left((1-c) s\right).
\label{z2}
\end{eqnarray}
Since $c < 1$, the first term in (\ref{z2}) vanishes in the limit $s\to \infty$. On the other hand, the asymptotic expansion $\hbox{Ei}(x) \sim e^x/x$ (for $x \to \infty$) indicates that the second term in (\ref{z2}) behaves like $1/s$.

Consequently, we expect two types of behavior of the anisotropy parameters at large times: exponential decay or power law. Our numerical studies indicate that the first possibility is realized for the cases where one or two anisotropy parameters are negative. Such cases correspond to mixed oblate-prolate or prolate configurations. The power law characterizes the systems which are initially oblate (the first anisotropy parameter is very large and positive, while the second parameter is positive or equals 0).

\end{document}